%% file: main.tex
\documentclass[runningheads]{llncs}
\usepackage{amsmath,graphicx}
\usepackage[hidelinks]{hyperref}
\usepackage[capitalise, noabbrev]{cleveref}
\usepackage{glossaries}
\usepackage{amsmath,amssymb,amsfonts}
\usepackage{algorithmic}
\usepackage{graphicx}
\usepackage{textcomp}
\usepackage{xcolor}
\usepackage{cite}
\usepackage{booktabs}
\usepackage{flushend}
\usepackage{enumitem}
\usepackage{siunitx}

\newacronym{tts}{TTS}{Text-To-Speech}
\newacronym{vc}{VC}{Voice Conversion}
\newacronym{asv}{ASV}{Automatic Speaker Verification}
\newacronym{la}{LA}{Logical Access}
\newacronym{fd}{FD}{First Digit}
\newacronym{stlt}{STLT}{short-term long-term}
\newacronym{svc}{SVC}{Support Vector Classifier}
\newacronym{rf}{RF}{Random Forest}
\newacronym{mfcc}{MFCC}{Mel-Frequency Cepstral Coefficient}
\newacronym{fc}{FC}{Fully Connected}
\newacronym{roc}{ROC}{Receiver Operating Characteristic}
\newacronym{auc}{AUC}{Area Under the Curve}

\begin{document}
 
\title{All-for-One and One-For-All: \\Deep learning-based feature fusion \\for Synthetic Speech Detection}
\author{Daniele Mari\inst{1}\orcidID{0000-0003-0727-3725} \thanks{This work was partially supported by the European Union under the Italian National Recovery and Resilience Plan (NRRP) of NextGenerationEU, partnership on “Telecommunications of the Future” (PE00000001 - program “RESTART”). Daniele Mari's activities were supported by Fondazione CaRiPaRo under the grants “Dottorati di Ricerca” 2021/2022.}
\and
Davide Salvi\inst{2}\orcidID{0000-0002-5163-3364} \and \\
Paolo Bestagini\inst{2}\orcidID{0000-0003-0406-0222} \and
Simone Milani\inst{1}\orcidID{0000-0001-8266-5839}}
\authorrunning{D. Mari et al.}
\titlerunning{Deep learning-based feature fusion for Synthetic Speech Detection}
\institute{University of Padova - Padova, Italy
\and
Politecnico di Milano - Milan, Italy\\
\email{\{daniele.mari, simone.milani\}@dei.unipd.it}\\ 
\email{\{davide.salvi, paolo.bestagini\}@polimi.it}
\\
}
\maketitle              
\begin{abstract}
Recent advances in deep learning and computer vision have made the synthesis and counterfeiting of multimedia content more accessible than ever, leading to possible threats and dangers from malicious users.
In the audio field, we are witnessing the growth of speech deepfake generation techniques, which solicit the development of synthetic speech detection algorithms to counter possible mischievous uses such as frauds or identity thefts.
In this paper, we consider three different feature sets proposed in the literature for the synthetic speech detection task and present a model that fuses them, achieving overall better performances with respect to the state-of-the-art solutions.
The system was tested on different scenarios and datasets to prove its robustness to anti-forensic attacks and its generalization capabilities.

\end{abstract}

\input{1_introduction}

\input{2_method}

\input{3_experimental_setup}
\input{4_results}
\input{5_conclusion}

\bibliographystyle{splncs04}
\bibliography{refs}

\end{document}

%% file: 1_introduction.tex
\section{Introduction}
\label{sec:intro}
\vskip -0.5ex
Recent developments in deep learning techniques and the increased availability of computational capabilities have made the generation and editing of multimedia content within everyone's reach.
While technological advances open the door to new possibilities~\cite{arstechnica}, their malicious utilization can produce severe consequences~\cite{bbcnews}.
An example of this phenomenon are deepfakes~\cite{nguyen2022deep}, synthetic multimedia content generated through deep learning techniques that depict individuals in actions and behaviors that do not belong to them. 
To address this rising problem, the scientific community has focused on developing detectors capable of discerning counterfeit material from authentic one~\cite{verdoliva2020media}.

In the audio field, two main categories of fake audio generation algorithms can be found, i.e., \gls{tts}~\cite{
wang2017tacotron, ren2019fastspeech} and \gls{vc}~\cite{tanaka2019atts2s, 
kaneko2019cyclegan} algorithms.
Several approaches have been proposed to detect such forged signals~\cite{zhang2017investigation}, ranging from methods that aim at detecting low-level artifacts~\cite{monteiro2020generalized, alzantot2019deep} to others that focus on more semantic aspects~\cite{conti2022deepfake, attorresi2022prosody}.
Among these, in~\cite{mari2022sound}, the errors in the \gls{fd} statistics w.r.t. the generalized Benford law are used as input to a \gls{rf} classifier to detect fake speech. The features are computed from the \gls{mfcc} of the analyzed audio tracks. Interestingly, the authors show that these are equally discriminative when computed on the whole audio track or only on the silent parts of the signal.
At the same time, in~\cite{borrelli2021synthetic}, the detection task is performed by exploiting a set of features inspired by the speech-processing literature, which are used as input of a supervised classifier.
These features aim to model speech as an auto-regressive process, simultaneously considering multiple auto-regressive orders.
Finally, the authors of~\cite{albadawy2019detecting} discriminate real and fake speech by leveraging higher-order bispectral correlations introduced by synthesis algorithms and not typically found in human speech.
The possibility of performing speech deepfake detection based on several methods is paramount in multimedia forensics.
Each method targets different footprints among the possible traces that can reveal a potential forging in forensic analysis. Combining multiple solutions makes creating a successful and undetectable threat increasingly difficult for attackers.
Furthermore, applying heterogeneous detection techniques increases the applicability of algorithms to different scenarios.

In this paper we consider three sets of features presented in state-of-the-art for synthetic speech detection and propose a model that fuses them to perform the same task.
The features we consider are: \gls{fd} features~\cite{mari2022sound}; \gls{stlt} features~\cite{borrelli2021synthetic}; bicoherence features~\cite{albadawy2019detecting}.
Then, the fusion process is performed following a deep-learning-based approach.
We consider these features as they analyze three different aspects of the audio signal: silence, speech, and bispectral correlations.
This is relevant since, as mentioned above, it can benefit the final detection performance of the model.
The proposed system outperforms the single models, as presented in the literature, and shows excellent robustness to anti-forensic attacks and good generalization capabilities on unseen datasets.

%% file: 2_method.tex
\section{Proposed system}
\label{sec:method}
\vskip -0.5ex
The proposed detector processes an audio signal $\mathbf{x}$
containing speech and classifies it with a label $y \in \{\text{REAL}, \text{FAKE}\}$, where REAL identifies authentic speech and FAKE indicates synthetically generated signals.
To do so, we consider three different feature sets proposed in the literature for synthetic speech detection and fuse them in a new system that performs the same task.
In particular, these correspond to the \gls{fd} features proposed in~\cite{mari2022sound}, the \gls{stlt} features presented in~\cite{borrelli2021synthetic} and the bicoherence features introduced in~\cite{albadawy2019detecting}.
We refer to these features as $\mathbf{f}_{\text{FD}}$, $\mathbf{f}_{\text{STLT}}$, and $\mathbf{f}_{\text{B}}$, respectively.
Regarding the \gls{fd} features $\mathbf{f}_{\text{FD}}$, we compute them only on the silenced parts of the signal under analysis, as it proved to be equally discriminative with respect to calculating them on the whole audio track~\cite{mari2022sound}.

We computed the feature sets $\mathbf{f}_{\text{FD}}$, $\mathbf{f}_{\text{STLT}}$ and $\mathbf{f}_{\text{B}}$ as proposed in the respective papers~\cite{mari2022sound, borrelli2021synthetic, albadawy2019detecting},  and we used them as inputs of the proposed models.
The dimensions of the three vectors are respectively $N_{\text{FD}}=$\num{416}, $N_{\text{STLT}}=$\num{800} and $N_{\text{B}}=$\num{8}. This could be a problem as, when using deep learning models, a high imbalance in the feature size can give the classifier a hard time learning how to assign the correct weight to its inputs.
To solve this issue, we projected the feature sets into different feature spaces using embedding extractors.
The resulting embeddings, that we call $\mathbf{e}_{\text{FD}}$, $\mathbf{e}_{\text{STLT}}$ and $\mathbf{e}_{\text{B}}$, contain \num{32}, \num{64} and \num{16} features respectively so that they have much more comparable size facilitating their integration.

The proposed system consists of four different blocks that need to be trained. Three are the \gls{fc} networks used to generate the embeddings $\mathbf{e}_{\text{FD}}$, $\mathbf{e}_{\text{STLT}}$, $\mathbf{e}_{\text{B}}$ from the input features, and one is the final model that performs the detection from the concatenation of the embeddings $\mathbf{e}_{\text{ALL}} = [\mathbf{e}_{\text{FD}}; \; \mathbf{e}_{\text{STLT}}; \; \mathbf{e}_{\text{B}} ].$
Table~\ref{tab:models} shows the architectures of the considered models, where the \gls{fc} layers indicated in bold are those used to extract the embeddings.
In all the networks we considered LeakyReLU as activation function and Softmax as output.

\begin{table}[!t]
\centering
\setlength{\tabcolsep}{12pt}
\caption{Architectures of the considered models. The \gls{fc} layers indicated in bold are those used to extract the embeddings.}
\label{tab:models}
\resizebox{.7\columnwidth}{!}{
\begin{tabular}{cccc}
\toprule
\textbf{\gls{fd} model} & \textbf{\gls{stlt} model} & \textbf{Bicoh. model} & \textbf{Fusion model} \\ \midrule \midrule
$\mathbf{f}_{\text{FD}}$: (416, 1) & $\mathbf{f}_{\text{STLT}}$: (800, 1) & $\mathbf{f}_{\text{B}}$: (8, 1) & $\mathbf{f}_{\text{ALL}}$: (112, 1) \\ \midrule
FC (416, 128) & FC (800, 512) & FC (8, 32) & FC (112, 32)    \\
Dropout (0.25) & Dropout (0.25) & Dropout (0.25) & Dropout (0.25)  \\
BatchNorm1D & BatchNorm1D & BatchNorm1D & BatchNorm1D \\ \midrule
FC (128, 64) & \textbf{FC (512, 64)} & \textbf{FC (32, 16)}     & FC (32, 2) \\
Dropout (0.25) & Dropout (0.25) & Dropout (0.25) &  \\
BatchNorm1D & BatchNorm1D & BatchNorm1D & \\ \midrule
\textbf{FC (64, 32)}     & FC (64, 2)               & FC (16, 2)               &                 \\
Dropout (0.25)           &                          &                 \\
BatchNorm1D              &                          &                          &                 \\ \midrule
FC (32, 2)               &                          &                          &                 \\ \midrule \midrule
$\mathbf{e}_{\text{FD}}$: (32, 1) & $\mathbf{e}_{\text{STLT}}$: (64, 1) & $\mathbf{e}_{\text{B}}$: (16, 1) & //    \\ \bottomrule
\end{tabular}}
\vspace{-1.5em}
\end{table}

\begin{figure}[!t]
    \centering
    \includegraphics[width=.6\columnwidth]{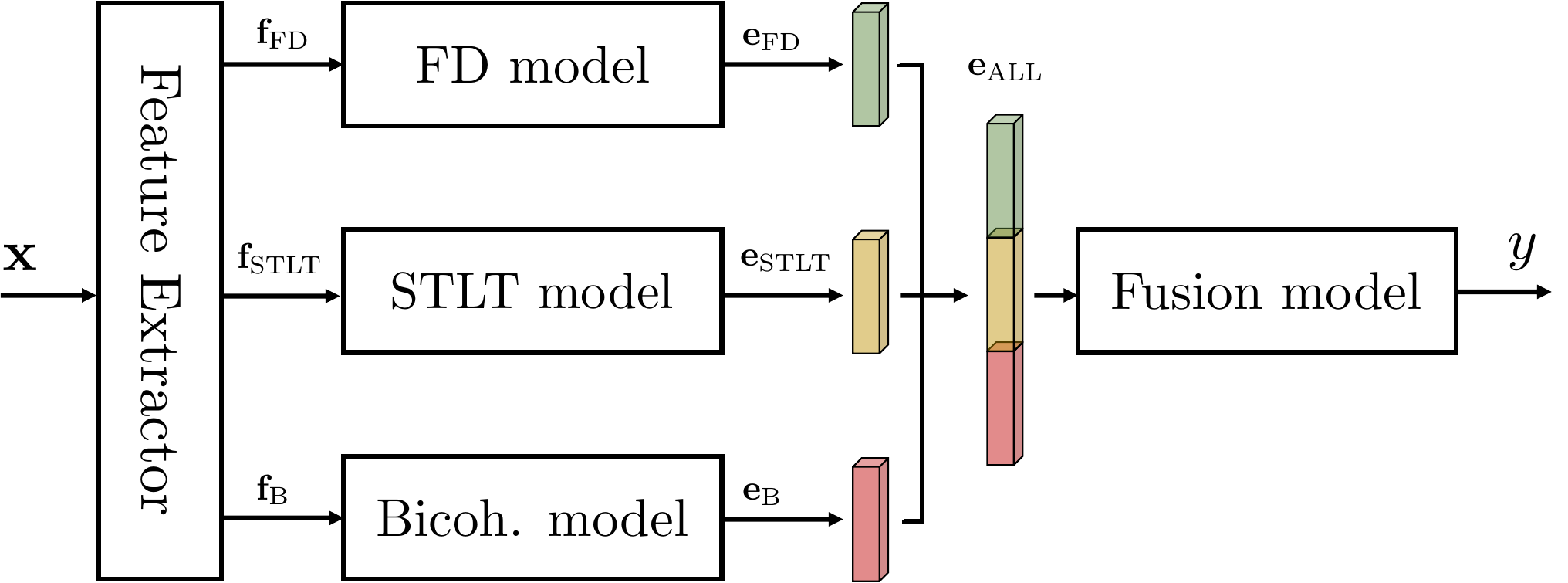}
    \caption{Pipeline of the proposed system.}
    \label{fig:pipeline}
    \vspace{-.5em}
\end{figure}

To summarize, the complete pipeline of the detection process is structured in three steps, as shown in Figure~\ref{fig:pipeline}.
First, the feature vectors $\mathbf{f}_{\text{FD}}$, $\mathbf{f}_{\text{STLT}}$, and $\mathbf{f}_{\text{B}}$ are extracted from the input speech signal under analysis $\mathbf{x}$.
Then, the three sets of features are given as input to the three separate \gls{fc} neural networks to perform dimensionality reduction and extract the embeddings $\mathbf{e}_{\text{FD}}$, $\mathbf{e}_{\text{STLT}}$ and $\mathbf{e}_{\text{B}}$.
Finally, a \gls{fc} network takes as input the concatenation $\mathbf{e}_{\text{ALL}}$ of the three generated embeddings and performs synthetic speech detection. The output of this model is $\hat{y}$, corresponding to the estimation of the class $y$ of the signal $\mathbf{x}$.

We decided to adopt a deep-learning-based approach to leverage most of the consistencies and inconsistencies within the analyzed feature sets.
The first three \gls{fc} models are used to perform a dimensionality reduction of the features, which helps create embeddings $\mathbf{e}$ that contain only the most crucial information of the corresponding feature vector $\mathbf{f}$.
In doing so, when we concatenate them, the features included in $\mathbf{e}_{\text{ALL}}$ are very informative for the task at hand and improve the detection capabilities of the system.

%% file: 3_experimental_setup.tex
\section{Experimental setup}
\label{sec:setup}
\vskip -0.5ex
In this section we provide the reader with some insights regarding the evaluation setup used to assess the performances of the proposed detector.
First, we describe the datasets considered for training and testing the system.
Then, we illustrate the training procedure used to train the \gls{fc} networks.
Finally, we define the anti-forensics attacks used to test the robustness of the model. 

\vspace{.5em}
\noindent
\textbf{Considered datasets.}
During all the experiments, we considered several datasets $\mathcal{D}$ to train and test the proposed synthetic speech detector. These include both real and deepfake speech samples for a total of more than \num{175000} audio tracks.
We used the following datasets to test the robustness of the proposed system in real-world conditions and verify that it is not overfitting to one set or domain.

\begin{itemize}[leftmargin=*]

    \item{\emph{ASVspoof 2019}} \cite{todisco2019asvspoof} is a speech audio dataset created to develop antispoofing techniques for automatic speaker verification. We consider its \gls{la} partition of the dataset. It contains both real and deepfake speech data generated with several techniques, including both \gls{tts} and \gls{vc} systems. In particular, the \textit{train} and \textit{dev} partitions include authentic signals along with speech samples generated with \num{6} different algorithms (named A01, A02, ..., A06), while the \textit{eval} partition comprises real signals and samples from \num{13} other algorithms (A07, ..., A19).
   
    \item{\emph{LJSpeech}} (LJS) \cite{ljspeech} contains audio clips of a single speaker reciting pieces from non-fiction books.

    \item{\emph{LibriSpeech}} (LS) \cite{panayotov2015librispeech} is an open-source dataset containing audio clips of authentic speech. From this corpus we considered the subset \textit{train-clean-100}.
    
    \item{\emph{Cloud2019}} corresponds to the dataset proposed in \cite{Lieto2019}. It includes tracks from different speakers generated considering several \gls{tts} cloud services: Amazon AWS Polly (PO), Google Cloud Standard (GS), Google Cloud WaveNet (GW), Microsoft Azure (AZ), and IBM Watson (WA).
   
    \item{\emph{VidTIMIT}} \cite{sanderson2009multi} comprises audio-video recordings of \num{43} people reciting short sentences and recorded in an office environment. We consider only the audio signals of these data since the videos are not related to the task at hand.

\end{itemize}

\vspace{.5em}
\noindent
\textbf{Training setup.}
We train the complete model in an end-to-end fashion, considering the whole system as a larger network, allowing it to identify the best configuration of parameters for the final classification.
Alternatively, we could train the single models individually for the speech deepfake detection task, performing a \textit{Late Fusion}~\cite{salvi2023robust} between the extracted embeddings.
However, we experienced that this second method led to poorer performance, so that all the results reported in the next section consider the end-to-end training strategy.

We trained the model for \num{100} epochs by monitoring the value of the validation loss, computed considering the Cross-Entropy function.
We assumed \num{10} epochs as early stopping, a batch size of \num{128}, and a learning rate $lr=$\num{e-4} appropriately reduced on plateaux.
Finally, the features we input to the model are normalized linearly between \num{0} and \num{1}.

Depending on the experiments, we trained the models on different datasets. 
In the first one, we only considered ASVspoof 2019 both in training and test. We did so to obtain results comparable with those of the papers which introduced the feature sets we are using \cite{mari2022sound, borrelli2021synthetic, albadawy2019detecting}.
We assumed the detectors of the single papers as baselines to benchmark our performance.
In this case, due to the high imbalance in the number of real and fake tracks contained in the ASVspoof dataset, we trained the models making sure we had the same number of samples of the two classes in each batch.

In the following experiments, on the other hand, we increased the number of real samples contained in the training set, considering both ASVspoof 2019 and LibriSpeech as training datasets.
This was done to balance the two classes and provide a broader distribution of real samples, improving the system's detection capabilities on unseen data.
In this case, however, we experienced unstable training due to the different distribution of the features extracted from the real samples of ASVspoof and LibriSpeech.
To address this issue, we decided to train the network by weighting the cost of each sample with weight	$$w(D, L) = \frac{1}{\lvert S(D, L)\rvert},$$
where $\lvert S(D, L)\rvert$ is the cardinality of the set of samples with label $L \in \{0, 1\}$ in dataset $D \in \{\text{ASVspoof }train, \text{Librispeech}\}$.
This proved effective in stabilizing the descent of the validation loss as it forced the classifier to learn meaningful features to discern real and fake samples instead of separating the distributions of the ASVspoof and Librispeech data.

To summarize, we have considered two different dataset configurations in training. In the first one we trained only on ASVspoof \textit{train} and \textit{dev} with batch balancing. We used this setup to benchmark the performances of our system with those of the reference papers~\cite{mari2022sound, borrelli2021synthetic, albadawy2019detecting}, which have been trained on these data.
In the second we trained the system ASVspoof and LibriSpeech with sample weighting. We considered this setup for the more challenging experiments, such as dataset generalization and anti-forensics attacks.

\vspace{.5em}
\noindent
\textbf{Anti-forensics attacks.}
The detection performance achieved by anti-deepfake systems may not be entirely significant in a real-world scenario. In fact, the quality of the ``in-the-wild'' signals degrades because of editing and formatting operations, thus compromising the accuracy of deepfake detection systems. This is the case of speech tracks found on the web or social platforms, often subject to compression and post-processing operations.

In this work, we test the robustness of the proposed detector for two of these operations, namely Gaussian noise injection and MP3 compression.
We do so since the presence of white noise could hide some artifacts introduced by synthetic audio generators, while compression is an operation often present in real-world scenarios.
We performed all post-processing operations using the Python \textit{audiomentations}~\cite{audiomentations}
library, 
considering three different noise levels with amplitude \textit{std} = [0.1, 0.01, 0.001] corresponding to \textit{SNR} = [2, 22, 42], and two different compression values (\textit{bitrate} [\si{kbit/s}] = [128, 32]).

%% file: 4_results.tex
\section{Results}
\label{sec:results}
\vspace{ -0.5ex}
In this section we analyze the features considered during the speech deepfake detection task and assess the performances of the proposed detector in different scenarios. 
We express these in terms of \gls{roc} curves, \gls{auc} and balanced accuracy.
Optimal performances are reached when \gls{auc} and balanced accuracy are equal to one.

\vspace{.2em}
\noindent
\textbf{Feature analysis.}
As a first experiment, we analyze the characteristics of each of the considered feature sets. 
We want to investigate how much the information they provide is correlated to avoid the computation of redundant data.
Since the proposed system performs a fusion of the three different sets of features, we want to ensure the content of the three is orthogonal so that we increase the amount of information fed to the model.
As previously mentioned, we use these features as they analyze three different aspects of a speech signal. $\mathbf{f}_{\text{FD}}$ contains information about silences, $\mathbf{f}_{\text{STLT}}$ models speech, while $\mathbf{f}_{\text{B}}$ exploits bispectral information.

\begin{figure}[!t]
    \centering
    \includegraphics[width=.5\columnwidth]{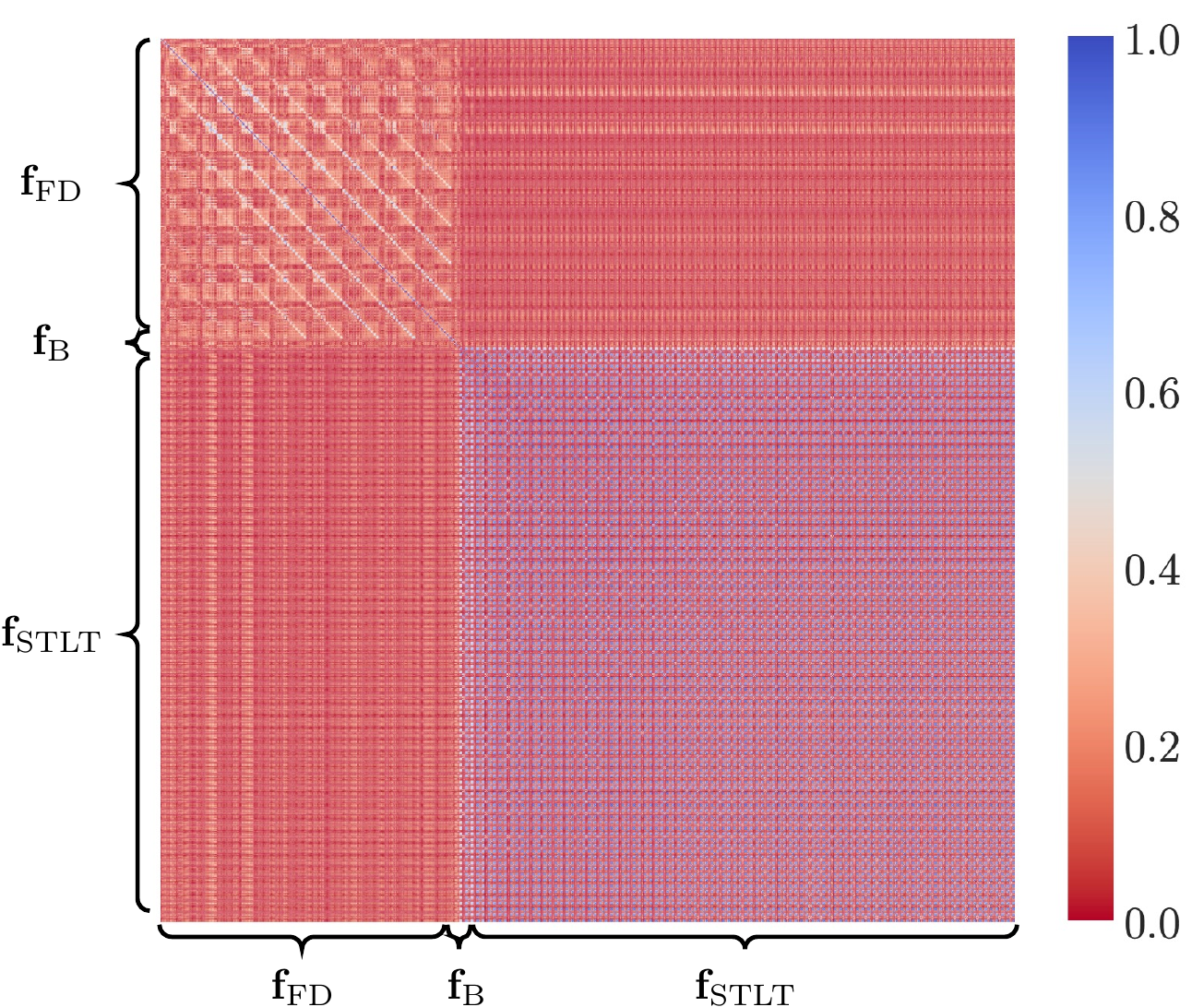}
    \vspace{-.5em}
    \caption{Cross-correlation matrix of the feature vectors $\mathbf{f}_{\text{FD}}, \mathbf{f}_{\text{STLT}}$ and $\mathbf{f}_{\text{B}}$ computed on the ASVspoof 2019 \textit{train} dataset. The absolute values of the coefficients are shown for visualization purposes.}
    \label{fig:correlation matrix}
    \vspace{-.5em}
\end{figure}

To measure the correlation between the three sets, we compute the Pearson coefficient for each pair of elements of the feature vectors.
The resulting matrix describes cross-correlations between different features and auto-correlations of each vector.
Figure~\ref{fig:correlation matrix} shows the absolute values of the results of this analysis computed on the ASVspoof 2019 \textit{train} dataset.
There, we can identify different rectangular regions for each feature vector. The bicoherence features are not clearly visible since they are much less numerous.
Although the single feature vectors present a quite high degree of internal correlation, the cross coefficients between them are low. This means that the three feature vectors do not strongly correlate and do not share much information.
This motivates the joint use of these features, increasing the model's detection accuracy and robustness and the use of \gls{fc} networks to perform dimensionality reduction and drop the redundant information within each feature set.

\begin{figure}[!t]
    \centering
    \includegraphics[width=0.5\columnwidth]{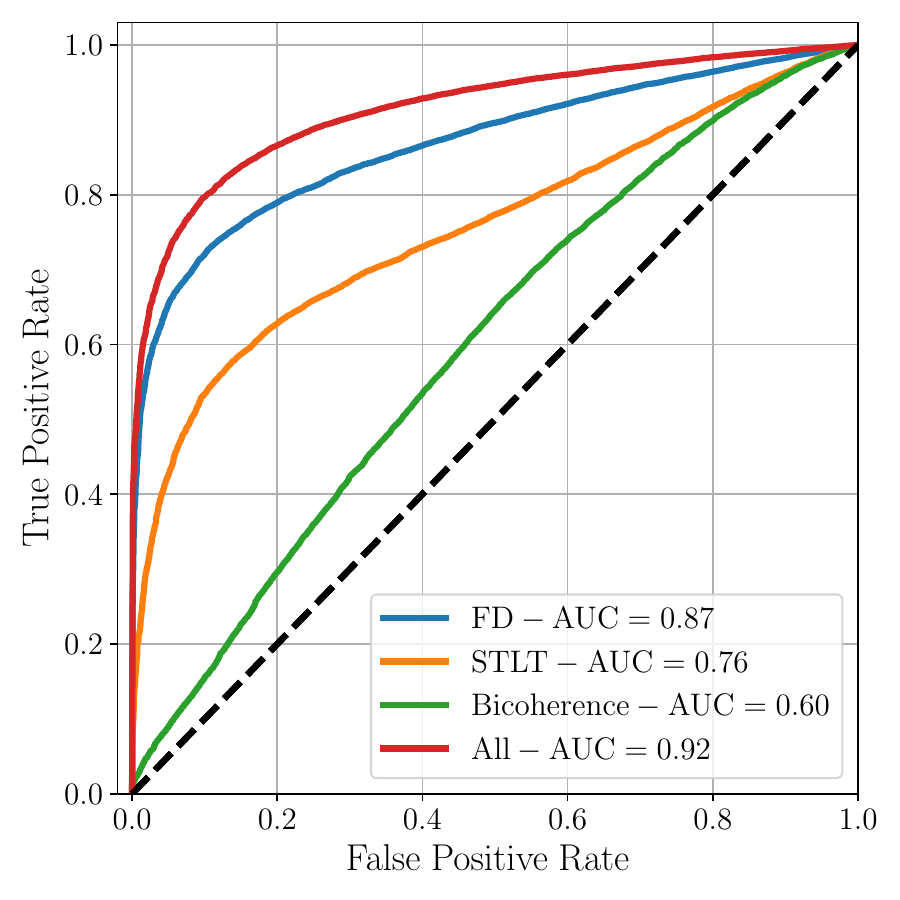}
    \vspace{-.5em}
    \caption{\gls{roc} curve on the ASVspoof 2019 \textit{eval} dataset of the four considered models.}
    \label{fig:roc}
    \vspace{-1em}
\end{figure}

\vspace{.2em}
\noindent
\textbf{Detection results.}
In this experiment we train and validate the proposed system respectively on the \textit{train} and \textit{dev} partitions of the ASVspoof 2019 dataset and test it on the \textit{eval} set.
We do the same with the single models that consider just one feature set at a time to verify that the fusion of the three actually improves the detection capabilities of the model.
Figure~\ref{fig:roc} shows the result of this analysis, where we can see that the fused model outperforms all the single ones and leads to better performances than the considered baselines.

Furthermore, we have also tried to see if a majority voting strategy between the predictions of the single classifiers was more effective than the proposed approach.
The end-to-end architecture shows superior performances that justify its use, with the balanced accuracy that increases by \num{6}\%.
This is probably because the end-to-end model can choose how to use and aggregate the information provided by the three feature sets, leading to better results.

\vspace{.5em}
\noindent
\textbf{Generalization results.}
We now assess the generalization capabilities of the proposed model by testing it on unseen data during training. 
This is an important aspect in multimedia forensics, where the developed detectors are often tested in conditions other than those considered during training and must be able to provide reliable predictions.
The datasets we consider in this experiment are Cloud19~\cite{Lieto2019}, LJSpeech~\cite{ljspeech}, and VidTIMIT~\cite{sanderson2009multi}. 
To improve the system performance in this scenario, we train it on both ASVspoof 2019 and LibriSpeech, following the same approach proposed in~\cite{conti2022deepfake}.
During training, we considered the weight-based strategy presented in the previous section.

\begin{figure}
     \vspace{-1em}
     \centering
     \includegraphics[width=0.52\columnwidth]{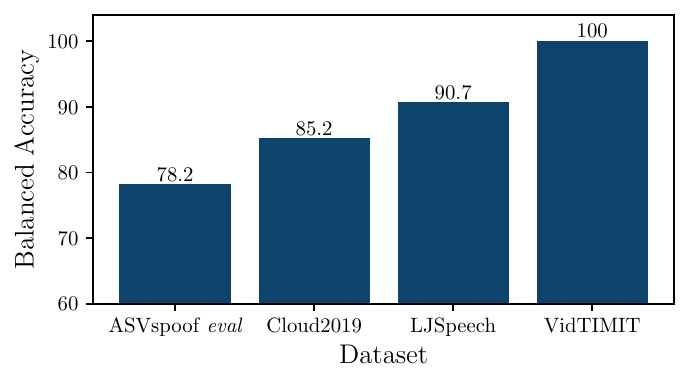}
     \vspace*{-1ex}
     \caption{Balanced accuracy values achieved by the proposed model on unseen datasets during training.}
     \label{fig:barplot}
     \vspace{-1.5em}
\end{figure}

Figure~\ref{fig:barplot} 
shows the results of this analysis, where the detector proves to have good generalization capabilities by scoring high accuracy values.
Furthermore, the joint training on ASVspoof and LibriSpeech has improved the accuracy values achieved on datasets never seen before.
As a drawback, the same strategy lowers the performance on ASVspoof 2019 \textit{eval}, with a balanced accuracy value that goes from \num{83.3} to \num{78.2} on the same dataset.
This is probably due to the fact that the system improves its generalization performance and becomes less prone to overfitting on ASVspoof.

\vspace{.5em}
\noindent
\textbf{Anti-forensics attacks.}
In this experiment we tested the robustness of the developed system to anti-forensics attacks.
Synthetic speech signals usually contain traces and artifacts left by the generators, which can be leveraged to discriminate them.
However, post-processing operations can corrupt or remove these, making the detection performance more challenging.
This is the case of media content circulating online, where the low quality and the post-processing applied to the signals make it challenging to classify them.
For this reason, being able to perform the deepfake detection task even on processed data is crucial.

We consider two anti-forensics attacks, Gaussian noise injection and MP3 compression, and evaluate how the detector performance is affected. As in the previous test, we consider the model trained on both ASVspoof 2019 and LibriSpeech.
A small change to the system must be implemented when we consider Gaussian noise injection.
Since the noise increases the power of the signal, it is unfeasible to detect the silenced regions (silence threshold needs to be automatically estimated as the analyst does not know the amount of added noise) and compute the \gls{fd} features on them. For this reason, we will consider \gls{fd} features computed on the whole signal. This should not affect the system final performance, as~\cite{mari2022sound} showed that the detector achieves equivalent scores when the \gls{fd} features are computed on the whole sample or the silenced parts only.

\begin{figure}[!t]
     \vspace{-1.5em}   
     \centering
     \includegraphics[width=0.49\columnwidth]{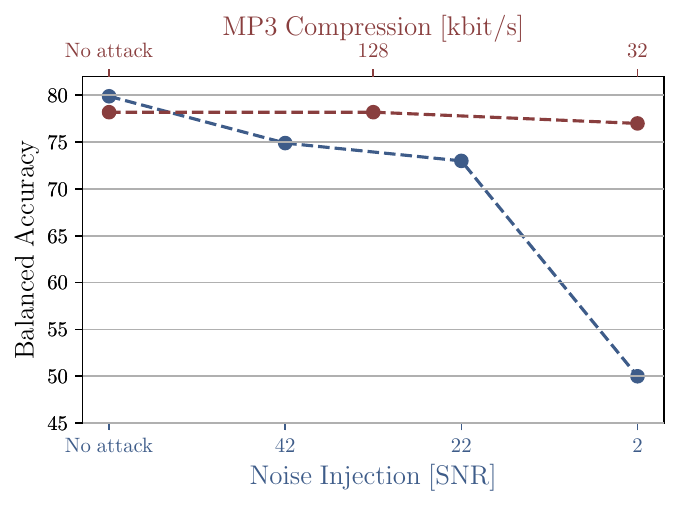}
     \vspace{-1em}
     \caption{Balanced accuracy values achieved by the proposed model when tested on data under anti-forensic attacks.}
     \label{fig:attacks}
     \vspace{-2em}
 \end{figure}

Figure~\ref{fig:attacks} 
shows the balanced accuracy values achieved on ASVspoof \textit{eval} in the considered cases.
These show that the proposed method is robust to MP3 compression since the accuracy does not seem to decrease even if we compress the sample considerably.
On the other hand, Gaussian noise appears to be more problematic as the prediction becomes almost random when injecting noise with a SNR value equal to \num{2}.

%% file: 5_conclusion.tex
\section{Conclusions}
\label{sec:conclusion}
\vskip -1ex
In this paper we proposed a synthetic speech detector based on a deep learning approach that fuses and harmonizes three feature sets proposed in the literature. The combination strategy is able to manage the different dimensionalities and mitigate some detection inefficiencies allowing to outperform networks trained independently on the single feature sets.
Furthermore, the proposed approach proved to generalize well on different datasets and achieved higher robustness with respect to anti-forensic attacks/conditions such as MP3 compression and Gaussian noise injection.
Future works will be dedicated to designing better feature fusion strategies by using more complex deep learning strategies, e.g., attention-based architectures that focus on specific audio excerpts that prove to be more discriminative information.